\documentclass[pra,letterpaper,preprint]{revtex4-2}
\usepackage[utf8x]{inputenc}
\usepackage{amsmath}\usepackage{xcolor}%
\usepackage{graphicx}
\usepackage{dcolumn}
\usepackage{bm}
\usepackage{subfigure}
\usepackage{textcomp}

\newcommand{\ket}[1]{\left| #1 \right\rangle}
\newcommand{\bra}[1]{\left\langle #1 \right|}

\newcommand{\abs}[1]{\left| #1 \right|}

\newcommand{\expn}[1]{{\rm e}^{#1}}

\newcommand{\dg}{{^{\dagger}}}

\newcommand{\ie}{\textit{i.e.,}}

\newcommand{\nnl}{\nonumber\\}

\begin{document}
	\title{Fundamental limits to the generation of highly displaced bright squeezed light using linear optics and parametric amplifiers}
	\author{Steve M. Young}%
	\email{syoung1@sandia.gov}
	\affiliation{Sandia National Laboratories, Albuquerque, New Mexico 87123
	}%
	\author{Daniel Soh}
	\email{danielsoh@arizona.edu}
	\affiliation{
		Wyant College of Optical Sciences, The University of Arizona, Tucson, Arizona 85721, USA
	}

	\date{\today}
	\begin{abstract}
	High quality squeezed light is an important resource for a variety of applications.  Multiple methods for generating squeezed light are known, having been demonstrated theoretically and experimentally. However, the effectiveness of these methods -- in particular, the inherent limitations to the signals that can be produced -- has received little consideration.  Here we present a comparative theoretical analysis for generating a highly-displaced high-brightness squeezed light from a linear optical method -- a beam-splitter mixing a squeezed vacuum and a strong coherent state -- and parametric amplification methods including an optical parametric oscillator, an optical parametric amplifier, and a dissipative optomechanical squeezer seeded with coherent states. We show that the quality of highly-displaced high-brightness squeeze states that can be generated using these methods is limited on a fundamental level by the physical mechanism utilized; across all methods there are significant tradeoffs between brightness, squeezing, and overall uncertainty.  We explore the nature and extent of these tradeoffs specific to each mechanism and identify the optimal operation modes for each, and provide an argument for why this type of tradeoff is unavoidable for parametric amplifier type squeezers.
\end{abstract}
	\maketitle
	
	\section{Introduction}

	Vacuum squeezed light has long been studied and used in various quantum applications \cite{SCHNABEL20171}. Particular interest lies in a squeezed light with high displacement and high brightness. This kind of quantum state becomes useful for various applications: for example, it can improve the measurement accuracy dramatically when the highly-displaced high-brightness phase-squeezed light is used for measuring optical phase delays \cite{zhuang2018distributed, guo2020distributed}. The phase measurement uncertainty is dictated by the phase variance of the state of the light, and highly-displaced high-brightness squeezed light is tremendously advantageous in reducing the phase uncertainty. It has been widely discussed that any metrology and sensing application would require both sensitivity and precision where the former can be improved by squeezed noise below the shot noise while the latter could only be achieved by high power probe light \cite{atkinson2021quantum}. Therefore, a highly-displaced high-brightness squeezed light will accomplish both the enhanced sensitivity and improved precision. A recent result of a super-resolution imaging using such highly-displaced high-brightness squeezed light attests the critical usage of such a quantum state of light \cite{soh2023label}. Alternatively, one can use a high-brightness amplitude-squeezed light to control an optically driven energy transition process \cite{kuzmich1997spin, milburn1984interaction}. This scenario becomes particularly relevant if the uncertainty of the final energy state after transition depends upon the uncertainty in the driving field's amplitude \cite{polzik1992atomic}. 
	
	\begin{figure}[htb!]
		\centering
		\subfigure[]{\includegraphics[width=0.24\textwidth]{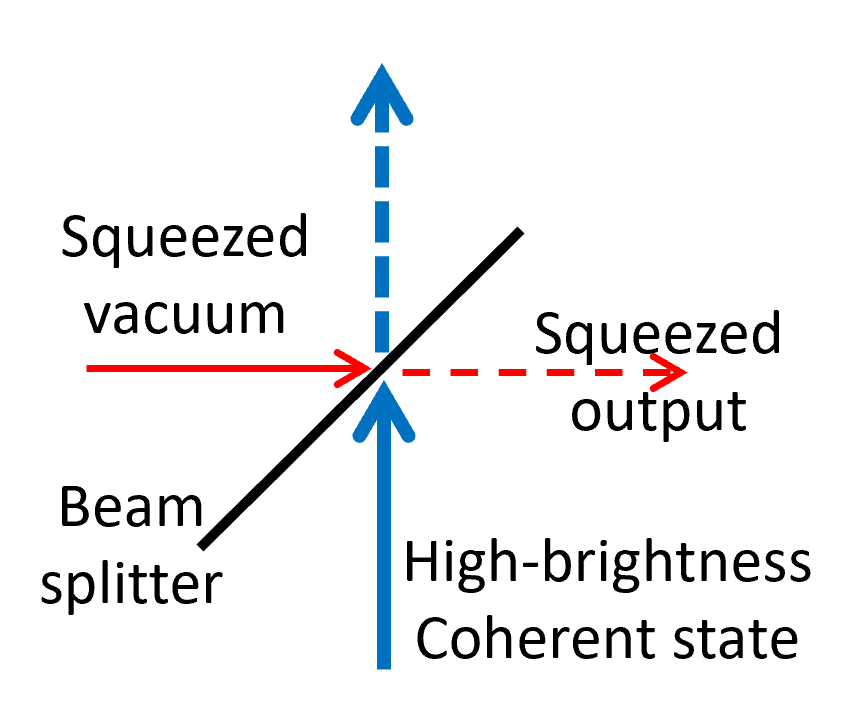}\label{fig:bs-m}} \\
		\subfigure[]{\includegraphics[width=0.31\textwidth]{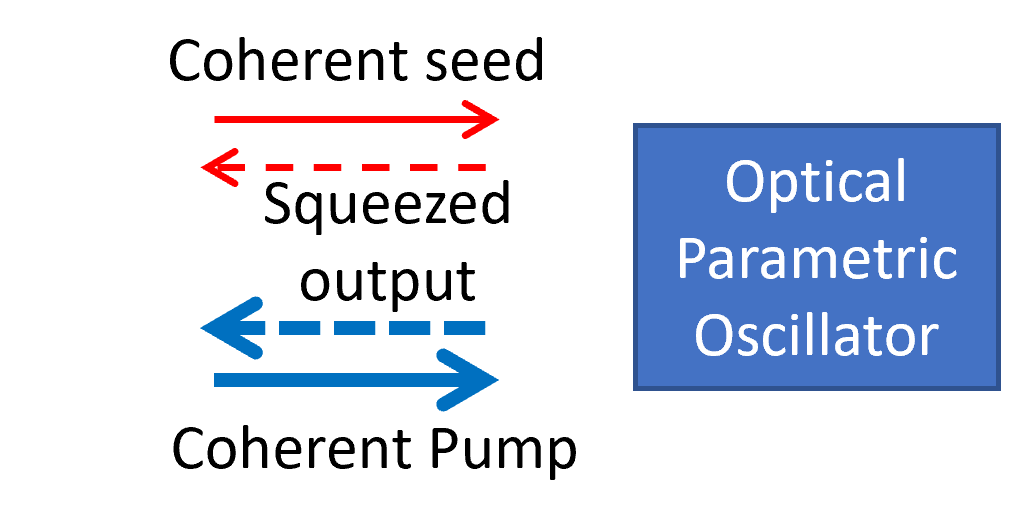}\label{fig:opo-m}} \\
		\subfigure[]{\includegraphics[width=0.31\textwidth]{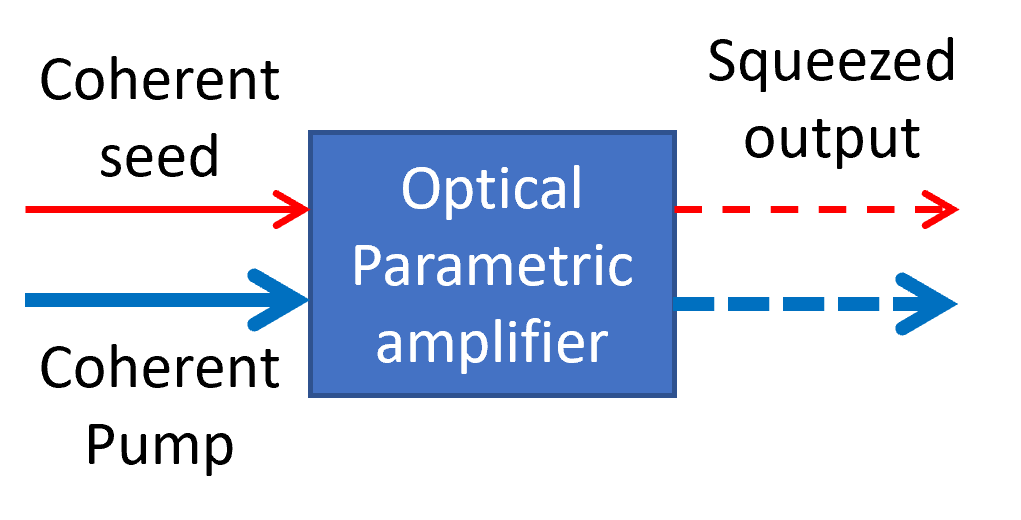}\label{fig:opa-m}} \\
		\subfigure[]{\includegraphics[width=0.36\textwidth]{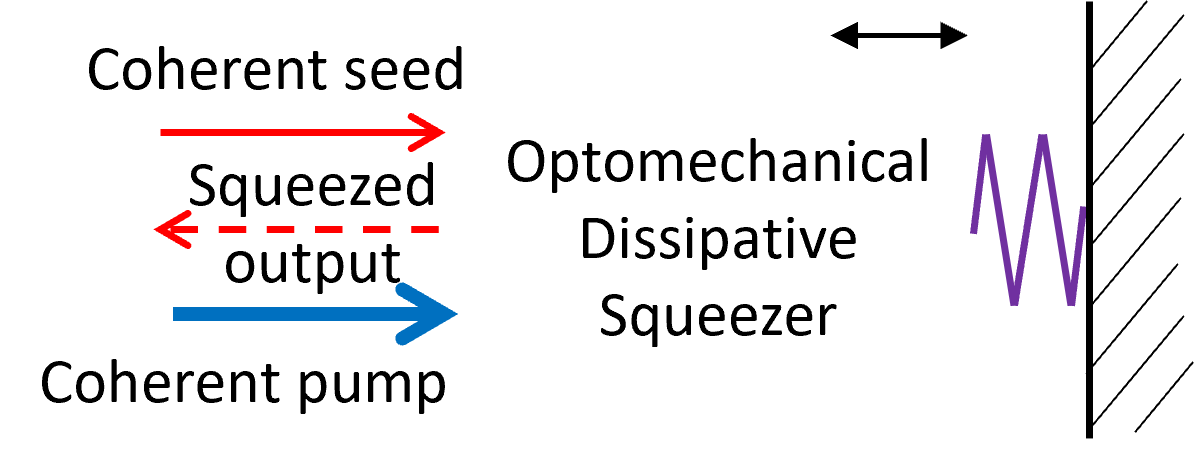}\label{fig:om-m}} \\
		\caption{Three distinct methods to create high-brightness squeezed light: \subref{fig:bs-m} beam-splitter mixing, \subref{fig:opo-m} seeded optical parametric oscillator, \subref{fig:opa-m} seeded optical parametric amplifier,and \subref{fig:om-m} seeded dissipative optomechanical light squeezing.} \label{fig:three-methods}
	\end{figure}

	Broadly, there are two possible approaches to generating highly-displaced high-brightness squeezed light.  The first is to generate a squeezed vacuum state using a given process and displace it through a beam splitter seeded with a strong coherent state of light.  The second is to seed the process for generating squeezed light so that it directly produces a highly-displaced squeezed signal rather than a squeezed vacuum.  However, in addition to engineering constraints, methods of squeezing face fundamental limits to the quality of the squeezed light that can be generated even for the rudimentary squeezed vacuum states.  One can expect that obtaining highly-displaced high-brightness squeezed states will come with additional restrictions.  
	
	In this work we compare methods to generate squeezed light with high displacement and high brightness and examine the \emph{fundamental} limits to the brightness and squeezing that can be obtained. First, we consider the case of beam-splitter mixing between a highly squeezed vacuum, provided by a method of choice, and high brightness coherent light.  Second, we consider the direct generation of highly-displaced bright squeezed light through seeding for the cases of optical parametric oscillators, optical parametric amplifiers, and dissipative optomechanical squeezers. We analyze the performance and the limitation of each method and evaluate the trade-offs among the brightness, the level of squeezing, and overall uncertainty of quadrature measurements. 
	
	\section{Beam-splitter mixing method}
	
	A straightforward method for creating a light state that has a reduced noise below the standard quantum limit (\ie shot noise) is to use a beam splitter and mix a squeezed vacuum and a high-brightness coherent state light (see Fig. \ref{fig:three-methods}(a)).  The two output modes from a beam splitter are generally entangled, unless the two input modes are coherent states. Therefore, the output that carries a reduced noise is indeed a mixed state after tracing out the other nearly coherent state light.  Still, the expected variance of the noise in the main squeezed output would be below the shot noise provided that the mixing ratio is sufficiently small. 
	
	The initial state of the two input modes in the tensor Hilbert space is 
	\begin{equation}
		\ket{\psi_i} = S_s(B)\ket{0}_s \otimes D_p (\mathcal{E}) \ket{0}_p,
	\end{equation}
	where the squeezing operator is given as \cite{caves1981quantum}
	\begin{equation}
		S_s(B) = \exp \left(\frac{1}{2} (B^* a_s^2 - B a_s^{\dagger 2}) \right),\label{eq:squeezing}
	\end{equation}
	$a, a^\dagger$ are the annihilation and the creation operators for the first input mode, and the displacement operator is
	\begin{equation}
		D_b (\mathcal{E}) = \exp \left( \mathcal{E} a_p^\dagger - \mathcal{E}^* a_p \right). \label{eq:displacement}
	\end{equation}
	Here, $\mathcal{E}$ is the complex amplitude of the displacement. Additionally, $\ket{0}_{s,p}$ represent a vacuum state in mode $s,p$, respectively. 
	
	The action of a beam splitter is best explained by a beam-splitter Hamiltonian given as
	\begin{equation}
		H_{\rm BS} = -\hbar \eta (a_s^\dagger a_p + a_s a_p^\dagger), \label{eq:bs-ham}
	\end{equation}
	where $\eta$ is the interaction (photon-exchange) strength. As shown in Appendix A, for $X$ displaced seed states in the limit of high pump displacement, the variance of quadrature measurements can be written as 
	
	\begin{align*}
		\left(\Delta X_s\right)^2 &= e^{-2 B} + \left(1-e^{-2 B}\right)\alpha^2,\\
		\left(\Delta P_s\right)^2 &= e^{2 B} + \left(1-e^{2 B}\right)\alpha^2.\\
	\end{align*}
	where $\alpha^2 (= \sin^2 \eta t)$, where $t$ is the time duration of beam splitter interaction, is the squared ratio of seed output displacement and pump input displacement. The squeezing, which is in amplitude for $B>0$ and in phase for $B<0$, is inversely related to the displacement squared of the output.  This result is quite physically intuitive as the two terms on the right-hand side explain the proportional noise addition between a squeezed vacuum and a displaced vacuum according to the beam-splitter ratio $\eta t$. Enhanced seed amplitude results from the mixing with the pump state, which contributes noise from the pump in addition to displacement from the pump. The overall uncertainty that we define as the product of the standard deviations in the measurements of both quadratures is
	\begin{flalign}
		\Delta X_s\Delta P_s=\sqrt{1+4\alpha^2\left(1-\alpha^2\right)\sinh^2\left(B\right)}.\label{eq:bs-u}
	\end{flalign}  
	reflecting the mixed state produced by the beam splitter.  Thus, in addition to accepting less squeezing as we increase the mixing ratio, we see that we also deviate increasingly from perfect squeezing that maintains the minimum overall uncertainty.  Allowing the beams to mix for a longer interaction time increases the output displacement but at the expense of squeezing magnitude and quality.

\begin{figure}[!tb]
		\centering
		\subfigure[]{\includegraphics[width=0.5\textwidth]{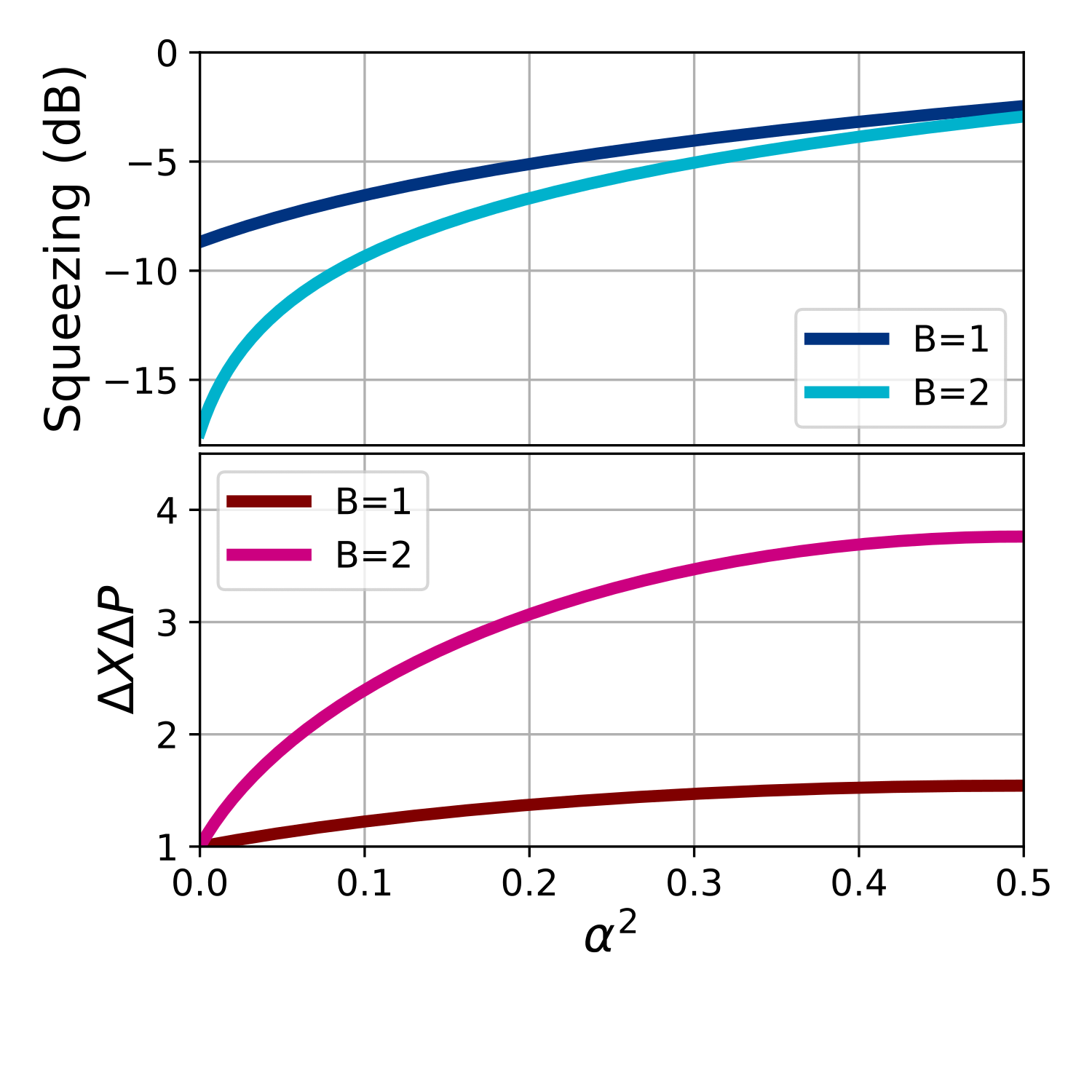}\label{fig:bs-squeeze_var}}		\subfigure[]{\includegraphics[width=0.5\textwidth]{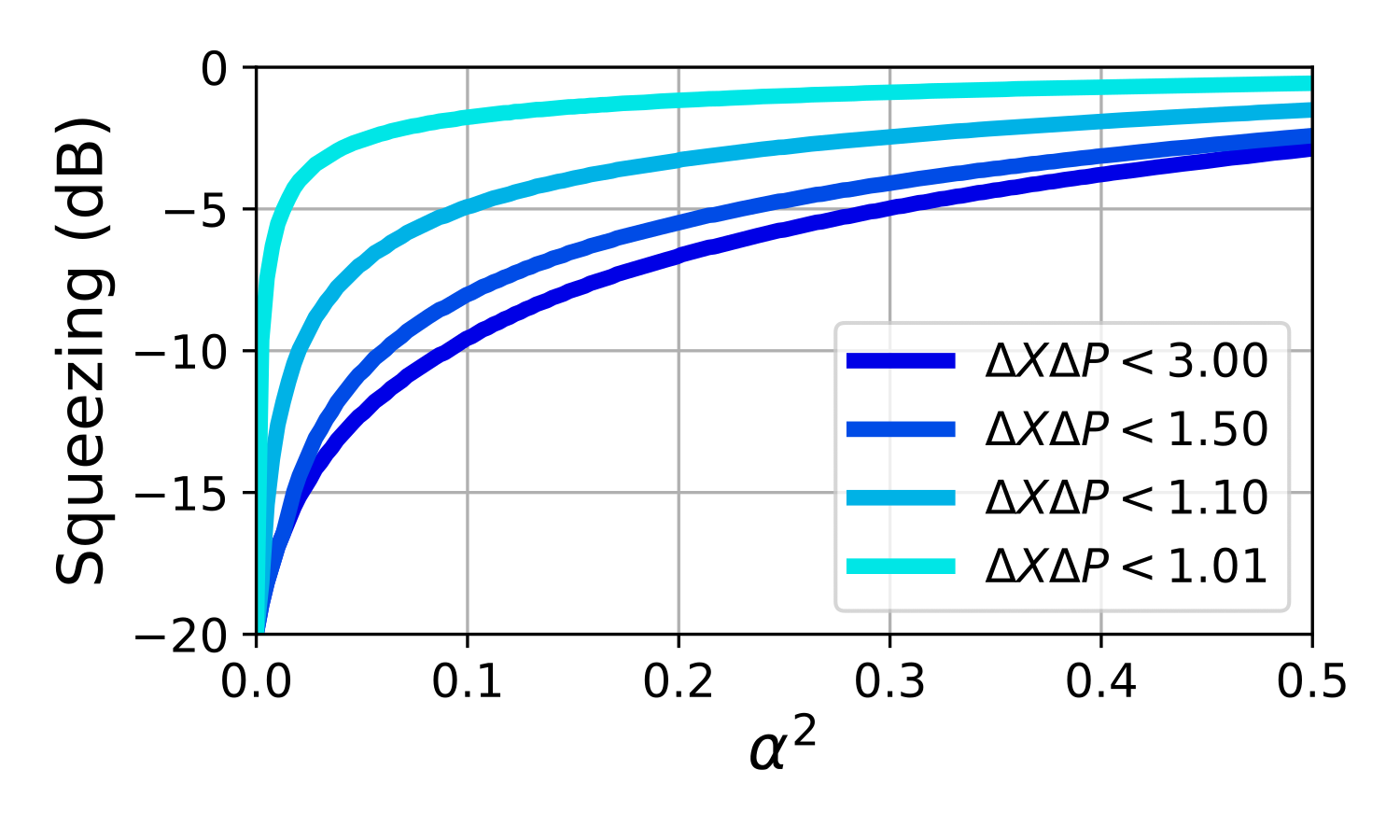}\label{fig:bs-squeeze_opt}}
		\caption{\subref{fig:bs-squeeze_var} The top panel shows the degradation of output squeezing factor as the relative squared output displacement $\alpha^2$ is increased for two different squeezing levels of the seed input. The bottom panel shows the accompanying increase in overall uncertainty. \subref{fig:bs-squeeze_opt} The best possible squeezing of the output obtainable with overall uncertainty below a given threshold.  The plots are representative both phase and amplitude squeezing, as the system operates symmetrically with respect to the two modes.} \label{fig:bs-squeeze}
	\end{figure}

	The consequences of this are shown in Fig.\ref{fig:bs-squeeze_var}.  First, we emphasize that there is a tradeoff between squeezing and output brightness, as the increase in noise variance is directly  proportionate to $\alpha^2$.  Additionally, the mixing of the pump signal impacts overall uncertainty.  If a weak (\ie low-displacement), but highly-squeezed output is desired it is optimal to use a highly squeezed vacuum input with low mixing rate (\ie small $\eta t$).  However, for brighter output states, using highly-squeezed vacuum input significantly increases overall uncertainty (Eq. \eqref{eq:bs-u});  minimizing overall uncertainty requires using a less squeezed vacuum state, which is somewhat counterintuitive. In Fig.\ref{fig:bs-squeeze_opt} we plot the maximum squeezing possible for a given $\alpha^2$, obtained by varying $\eta t$ and $B$ constrained by constrained $\alpha^2$, allowing for the visualization of both tradeoffs together. Ultimately, there is an unavoidable tradeoff between brightness and squeezing of the output inherent in the beam splitter mechanism. 
	
	\section{Optical Parametric Oscillator/Amplifier}
	We next analyze and predict the performance of the second method - a seeded degenerate optical parametric amplifier or oscillator (Fig. \ref{fig:three-methods} (b,c)). In these systems a nonlinear medium couples pump signal to a half-frequency seed signal via the interaction Hamiltonian
	\begin{align}
		H_{\rm OP} &= \hbar \omega a_s^\dagger a_s + \hbar 2 \omega a_p^\dagger a_p  +i \hbar (g/2) (a_s^{\dagger 2} a_p - a_s^2 a_p^\dagger), \label{eq:OPO-hamiltonian}
	\end{align}
	where $a_p$ and $a_s$ are the field operators for the pump and seed signals, respectively, $g$ is the field coupling, and $\omega$ is the seed signal frequency.  At present it is the method of choice for generating squeezed light, with greater than $10$~dB squeeze factors obtainable \cite{Vahlbruch2008,Eberle2010,Stefszky2012,Vahlbruch2016}.
	
	Depending on the relative phase of the pump ($p$) and seed ($s$) signals, the modified seed signal is either amplified or deamplified by the interaction.  Here we will take the pump and seed inputs to be $X$ displaced, with the operating mode depending on the relative sign of the pump displacement with respect to the seed displacement. Additionally, the output experiences phase or amplitude squeezing, respectively.  In the case of a parametric amplifier the signals are fed directly through the medium, whereas in the case of a parametric oscillator the medium in placed inside of a cavity, into which the signals enter and interact.

	We first consider the case of the parametric oscillator where input and output occur on a single side of the cavity.	This was first analyzed in \cite{collet1984opo} ignoring pump field noise, where it was found that the squeezing depends on the ratio $\frac{g\langle a_p\rangle}{\kappa/2}$ and that maximally phase-squeezed output is possible when $\frac{g\langle a_p\rangle}{\kappa/2}=1$.  However, this result was for the non-seeded case so that the squeezed output state is a vacuum. It was shown later \cite{lariontsev2002characteristics} that the presence of a seed signal requires explicit consideration of the noise present in the pump signal and significantly alters the squeezing that can be obtained.  In particular, unlike the beam-splitter case the choice of squeezing axis dramatically impacts the squeezing performance, which is significantly impaired by high seed power.  
	
	The equations of motions and their solutions can be found in Appendix B.  The results are easiest to understand in the limit of small $\mathcal{E}_s^{\rm in}$ (which is the input field displacement.) Perturbatively expanding around the internal seed displacement $\mathcal{E}_s^{\rm in}=0$, the relative squared output displacement is
	\begin{flalign*}
		\alpha^2=\left(\frac{\mathcal{E}_s^{\rm out}}{\mathcal{E}_p^{\rm in}}\right)^2=\left(\frac{1+\mathcal{C}_0}{1-\mathcal{C}_0}\frac{\mathcal{E}_s^{\rm in}}{\mathcal{E}_p^{\rm in}}\right)^2,
		\end{flalign*}
	where $\mathcal{E}$ are the input and output field displacements for the pump and seed, as indicated by the subscripts and superscripts. Here, $\mathcal{C}_0=\frac{gA_p^0}{\kappa/2}$, where $A_p^0$ is the intracavity pump amplitude for $\mathcal{E}_s^{\rm in}=0$, and $\kappa$ is the cavity dissipation rate. The variances of the output quadratures (Eq.\eqref{eq:opo-quad}) are
	\begin{flalign*}
		\left(\Delta X_s\right)^2&=\left(\frac{\frac{3\mathcal{C}_0^2}{2\left(1+\mathcal{C}_0\right)^2}\alpha^2-\left(1+\mathcal{C}_0\right)}{\frac{3\mathcal{C}_0^2}{2\left(1+\mathcal{C}_0\right)^2}\alpha^2+\left(1-\mathcal{C}_0\right)}\right)^2+\frac{\frac{4\mathcal{C}_0^2}{\left(1+\mathcal{C}_0\right)^2}}{\left(\frac{3\mathcal{C}_0^2}{2\left(1+\mathcal{C}_0\right)^2}\alpha^2+\left(1-\mathcal{C}_0\right)\right)^2}\alpha^2, \\
		\left(\Delta P_s\right)^2&=\left(\frac{\frac{\mathcal{C}_0^2}{2\left(1+\mathcal{C}_0\right)^2}\alpha^2-\left(1-\mathcal{C}_0\right)}{\frac{\mathcal{C}_0^2}{2\left(1+\mathcal{C}_0\right)^2}\alpha^2+\left(1+\mathcal{C}_0\right)}\right)^2+\frac{\frac{4\mathcal{C}_0^2}{\left(1+\mathcal{C}_0\right)^2}}{\left(\frac{\mathcal{C}_0^2}{2\left(1+\mathcal{C}_0\right)^2}\alpha^2+\left(1+\mathcal{C}_0\right)\right)^2}\alpha^2.
	\end{flalign*}
	As with the beam splitter case, we have written the quadrature variances in terms of $\alpha^2$ to highlight the connection to the noise.
	
\begin{figure*}[!tb]
		\centering
		\subfigure[Phase squeezing]{\includegraphics[width=0.45\textwidth]{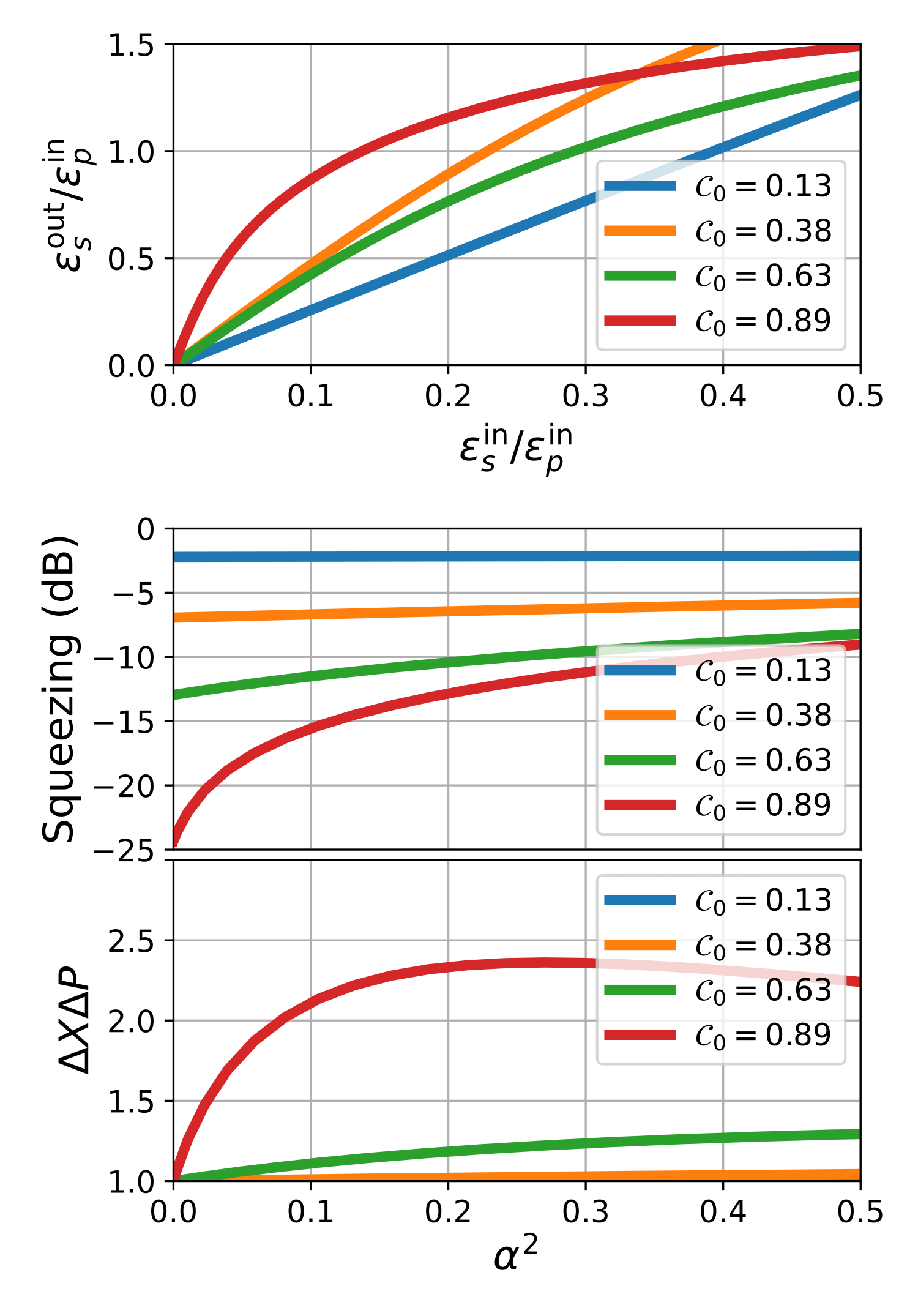}\label{fig:opo-psqueeze_var}}		
		\subfigure[Amplitude squeezing]{\includegraphics[width=0.45\textwidth]{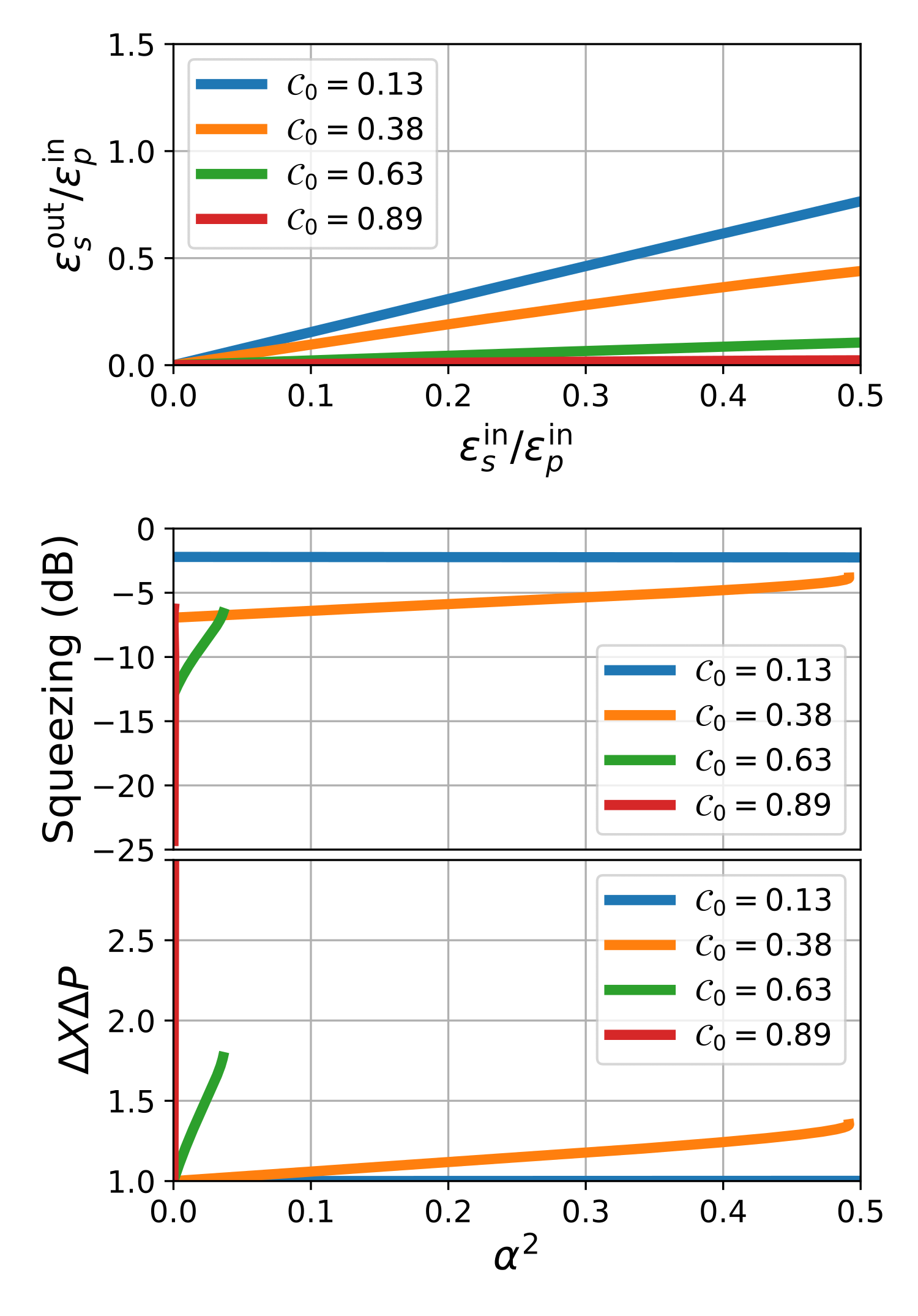}\label{fig:opo-xsqueeze_var}}
		\caption{Characteristics of the output signal for \subref{fig:opo-psqueeze_var} phase-squeezing and \subref{fig:opo-xsqueeze_var} amplitude-squeezing. In the top panels, the amplitude of the seed output relative to the pump input is plotted as a function of time for different values of $\mathcal{C}_0=\frac{gA_p^0}{\kappa/2}$. In the center and bottom panels, the corresponding squeezing factor and overall uncertainty of the output are plotted against relative squared output displacement $\alpha^2$.     For amplitude squeezing, the squeezing and uncertainty curves are cutoff where the relationship to target output power becomes nonmonotonic due to the high seed input required to overcome deamplification.
  } \label{fig:opo-squeeze_var}
	\end{figure*}

	The first terms on the RHS can be attributed to noise originating from the seed input, while the second terms are due to the noise from the pump input. We can see that for vacuum output we recover the original expressions in \cite{collet1984opo} that allow for arbitrarily strong squeezing, while increasing the desired output through a displaced seed input adds noise to both quadratures proportional to $\alpha^2$, necessarily compromising the squeezing and overall uncertainty.  To see this quantitatively, results calculated using the exact expressions Eqs. \eqref{eq:opo-quad} and \eqref{eq:opo-output} are shown in Fig. \ref{fig:opo-squeeze_var}.

	When $\mathcal{E}_p$ is negative and $\mathcal{E}_s$ is positive, the seed signal is amplified and the output signal is phase squeezed.  This case is shown in Fig.~\ref{fig:opo-psqueeze_var} for different values of $\frac{gA_p}{\kappa/2}$.  The amplification of the signal aids in the generation of a bright signal for a given level of squeezing; the parameter regime that results in the strongest amplification coincides with the greatest level squeezing.  For any desired brightness the greatest level of squeezing is obtained for  $\frac{gA_p}{\kappa/2}$ very close to 1.  However, this comes at a cost: the overall uncertainty is increased dramatically in this regime.  
	
	When $\mathcal{E}_p$ and $\mathcal{E}_s$ are both positive (Fig~\ref{fig:opo-xsqueeze_var}), the seed signal is deamplified and the output signal is amplitude squeezed. In contrast to the previous case, the high values of  $\frac{gA_p}{\kappa/2}$ yielding greater squeezing suppress the brightness of the output.  This leads to a more complicated picture of squeezing vs. power; depending on the desired power, the greatest squeezing will be associated with different values of  $\frac{gA_p}{\kappa/2}$.  The situation is also much worse in terms of overall uncertainty.   
	
	\begin{figure}[!tb]
		\centering
		\subfigure[Phase squeezing]{\includegraphics[width=0.45\textwidth]{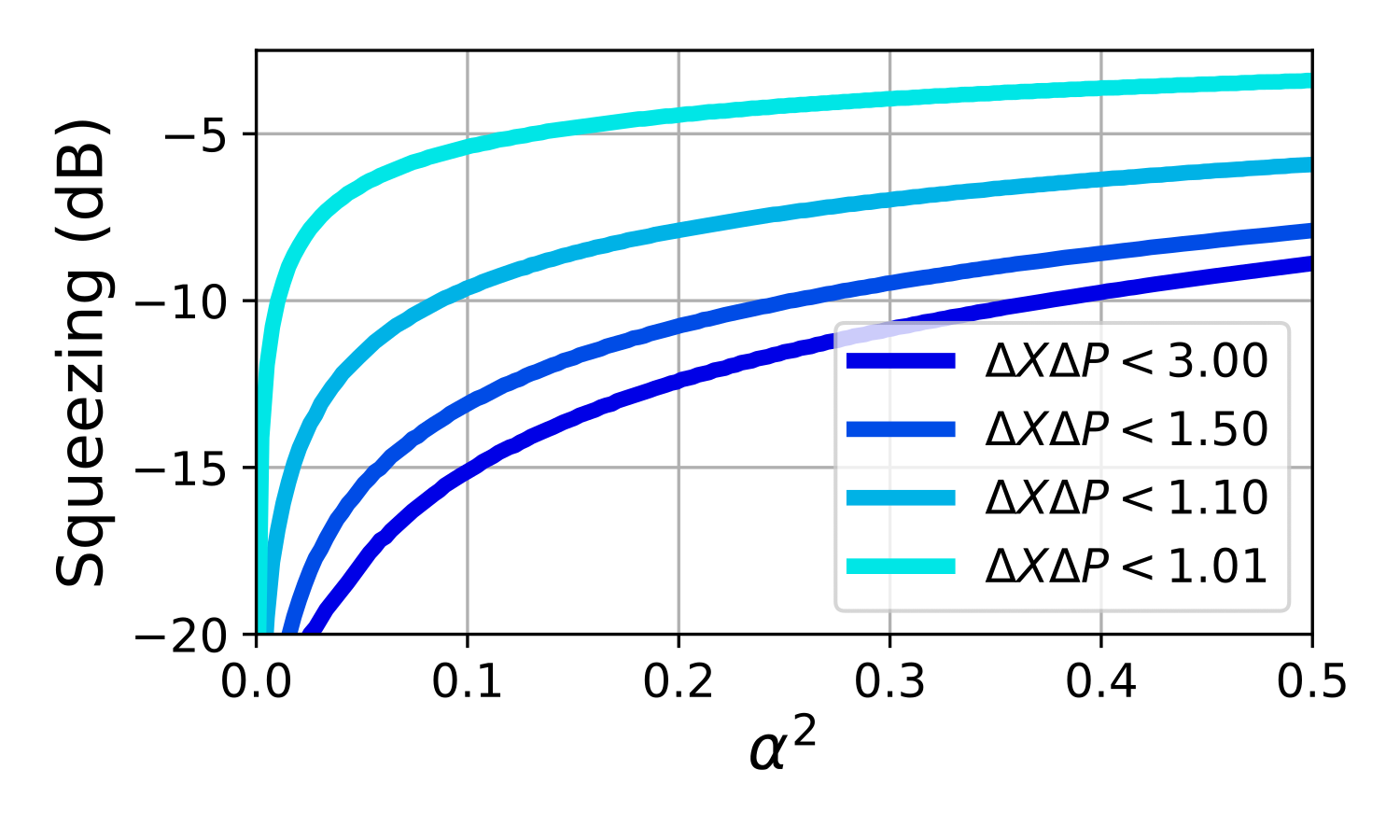}\label{fig:opo-psqueeze_opt}}		
        \subfigure[Amplitude squeezing]{\includegraphics[width=0.45\textwidth]{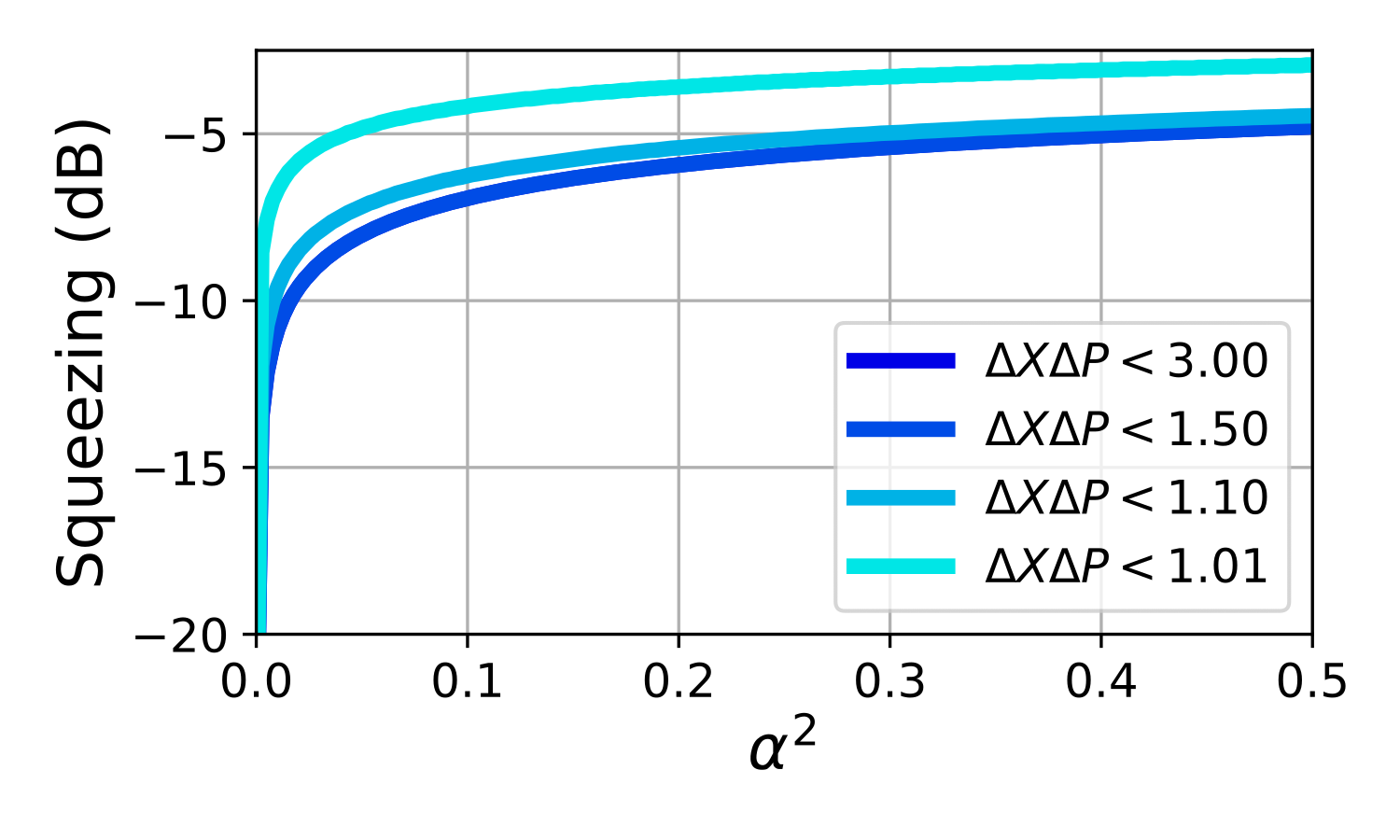}\label{fig:opo-xsqueeze_opt}}
		\caption{The optimal squeezing attainable for a parametric oscillator operating in the \subref{fig:opo-psqueeze_opt} amplifying, phase squeezing and \subref{fig:opo-xsqueeze_opt} deamplifying, amplitude squeezing regimes.  Each curve shows the best squeezing possible while maintaining total uncertainty below a given threshold.} \label{fig:opo-squeeze_opt}
	\end{figure}
	
	As in the beam-splitting case, we can summarize these results by plotting the optimal squeezing vs. $\alpha^2$ in Fig.\ref{fig:opo-squeeze_opt} for different thresholds of overall uncertainty.  As anticipated, generation of phase-squeezed light can outperform the beam splitter while generation of amplitude-squeezed light underperforms.  In this latter case, the optimal strategy would be to generate a squeezed vacuum state and displace it via a beam splitter.

\begin{figure*}[!tb]
	\centering
	\subfigure[Phase squeezing]{\includegraphics[width=0.45\textwidth]{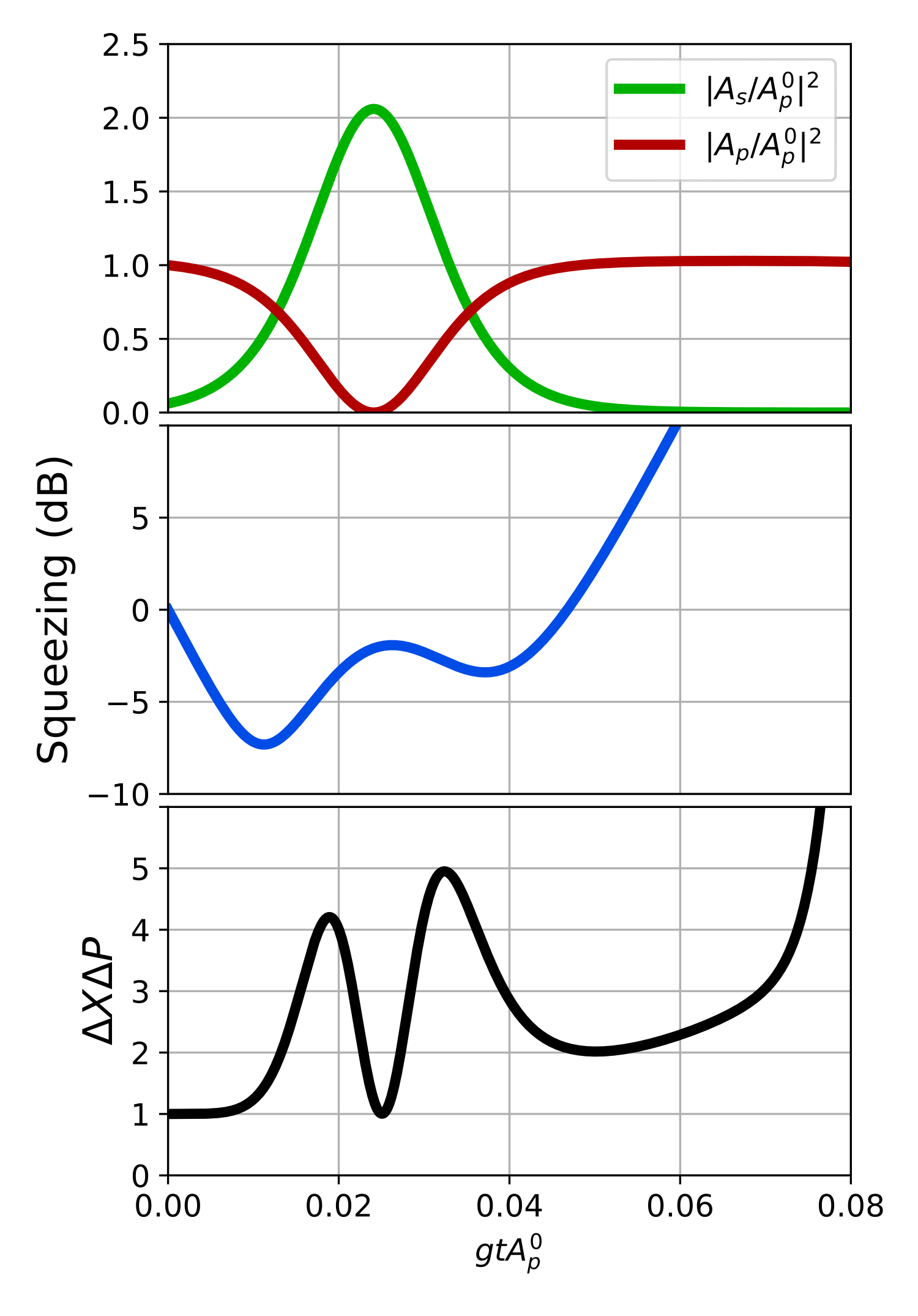}\label{fig:opa-psqueeze_var}}		
	\subfigure[Amplitude squeezing]{\includegraphics[width=0.45\textwidth]{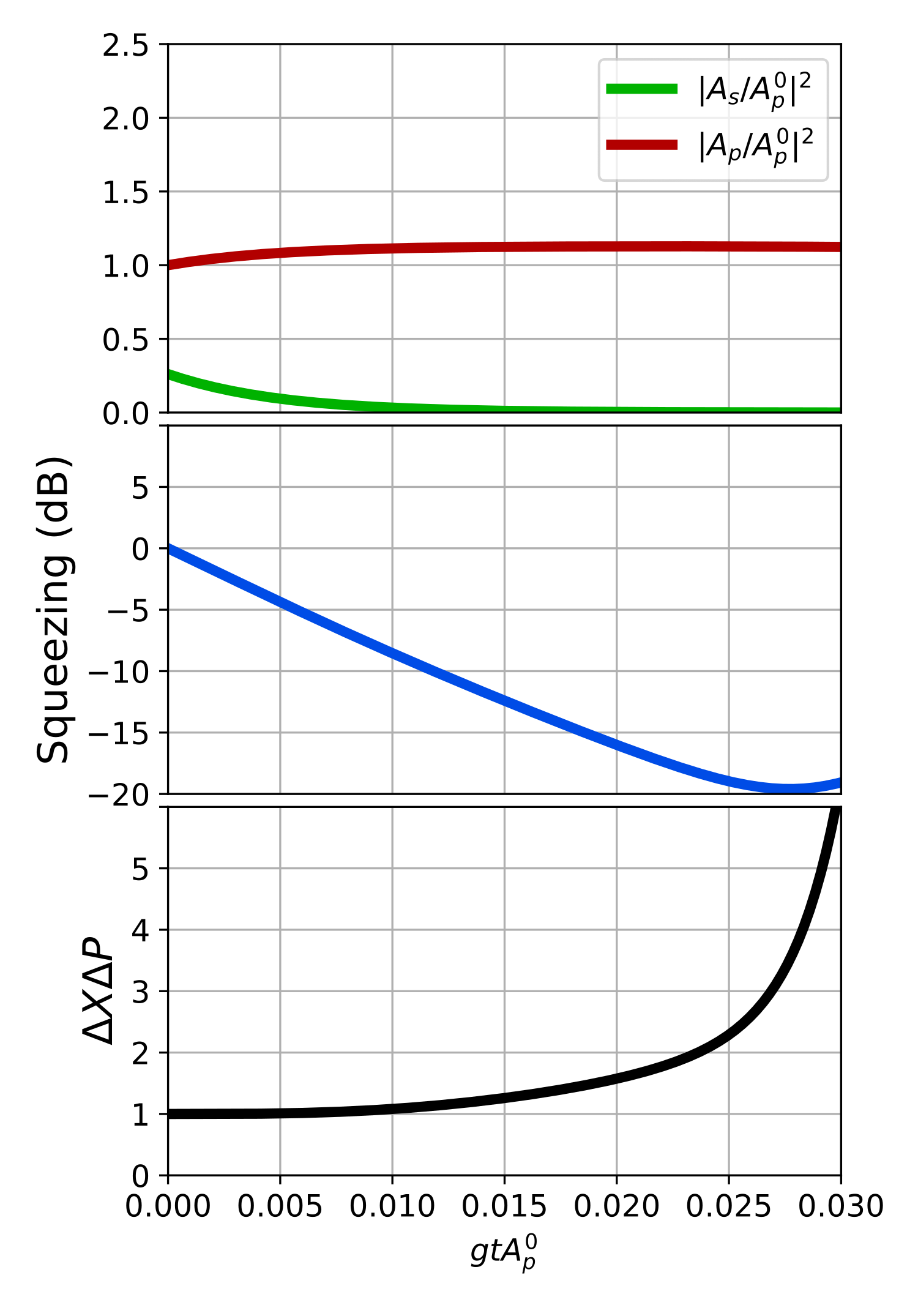}\label{fig:opa-xsqueeze_var}}
	\caption{Dynamics of the squared field amplitudes (top), squeezing (middle), and overall uncertainty (bottom) for \subref{fig:opa-psqueeze_var} amplification/phase-squeezing and \subref{fig:opa-xsqueeze_var} deamplification/amplitude-squeezing. } \label{fig:opa-squeeze_var}
\end{figure*}

	Turning our attention to the case of the parametric amplifier, we must consider non-steady-state dynamics as the pump and seed fields evolve in the medium.  This duration becomes the relevant timescale in place of $2/\kappa$ in the oscillator case.  The equations of motion are presented in Appendix C.  To compute the squeezing we integrate equations for the noise operators in Eq.\eqref{eq:opa_eom} numerically.  Fig.~\ref{fig:opa-squeeze_var} shows example trajectories of relevant quantities for both the amplifying, phase squeezing regime and the deamplifying, amplitude squeezing regime.  The top panel shows the field amplitudes.  It is clear that the amplitudes of both fields change dramatically, even for minimal seeding.  Importantly, at long enough times the seed amplitude always vanishes. 

	In the deamplifying, amplitude-squeezing case, the dynamics are straightforward.  The seed is depleted while the squeezing increases until the impact of pump noise mixing causes squeezing to degrade along with increasing overall uncertainty. In the amplifying, amplitude-squeezing case the seed amplitude increases as the pump is depleted. Then, at the point where the pump can no longer be depleted, the pump amplitude changes its course to increase and the system enters the deamplifying regime, under which the seed is depleted and the similar dynamics to the deamplifying case prevail.   In middle and bottom panels we show the corresponding trajectories of squeezing and overall uncertainty.  As with amplitude, the squeezing changes dramatically over time.  Initially the squeezing increases linearly. However, the rate of squeezing degrades as the pump depletes and the pump noise and seed noise mix. This reverses when the mixing-in of the pump noise, which includes a substantial component of mixed-in seed noise of opposite sign, exceeds the squeezing.  Once the system enters the deamplification regime, the long term effect is phase squeezing and rapidly increasing overall uncertainty as the seed and pump noise continue to mix. This is distinctively different from the amplitude-squeezing only case where such noise mixing does occur only at the excessively squeezing stage. Importantly, the maximum squeezing occurs early in evolution, well before the timing of the local extrema of the amplitudes.
	
		\begin{figure}[!tb]
		\centering
		\subfigure[]{\includegraphics[width=0.45\textwidth]{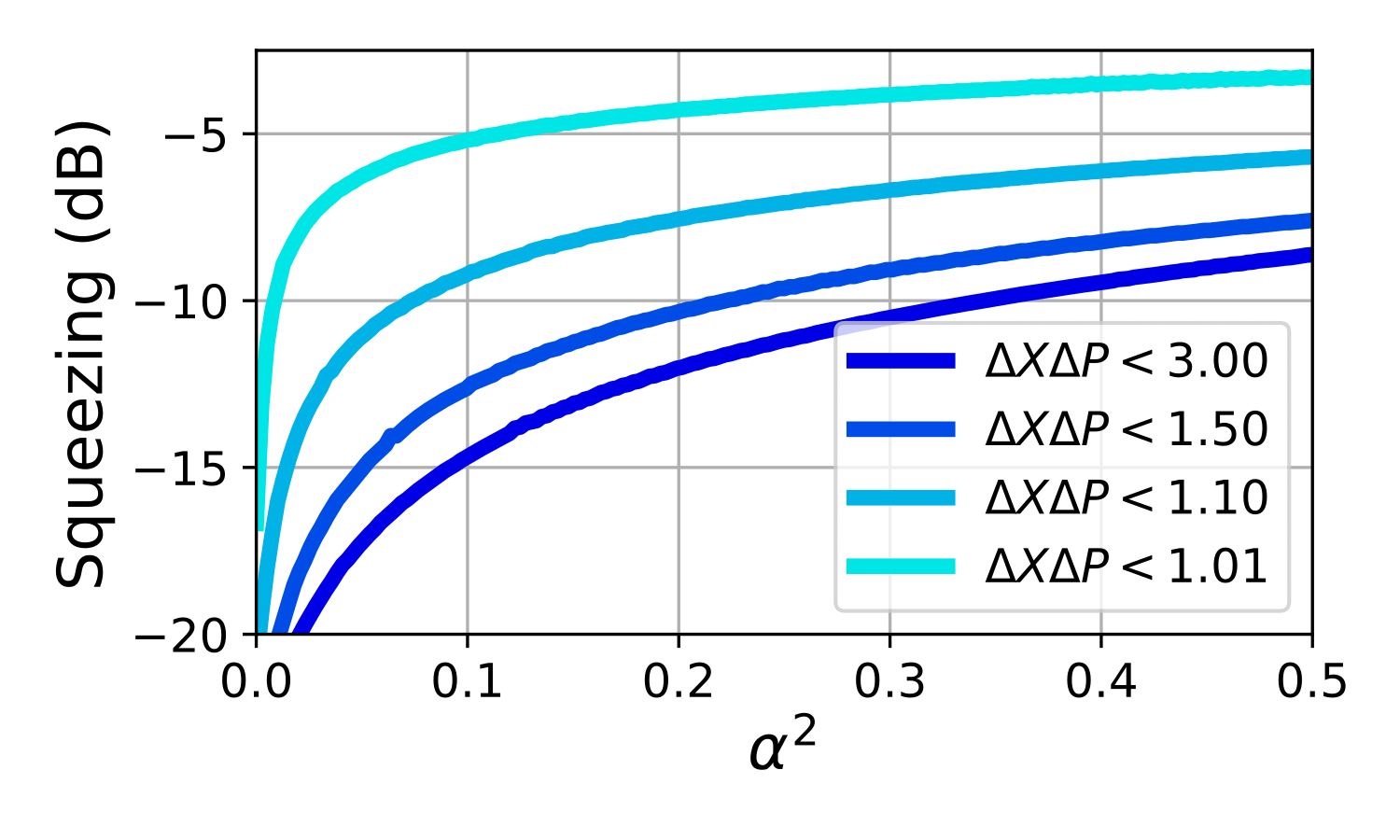}\label{fig:opa-psqueeze_opt}}		\subfigure[]{\includegraphics[width=0.45\textwidth]{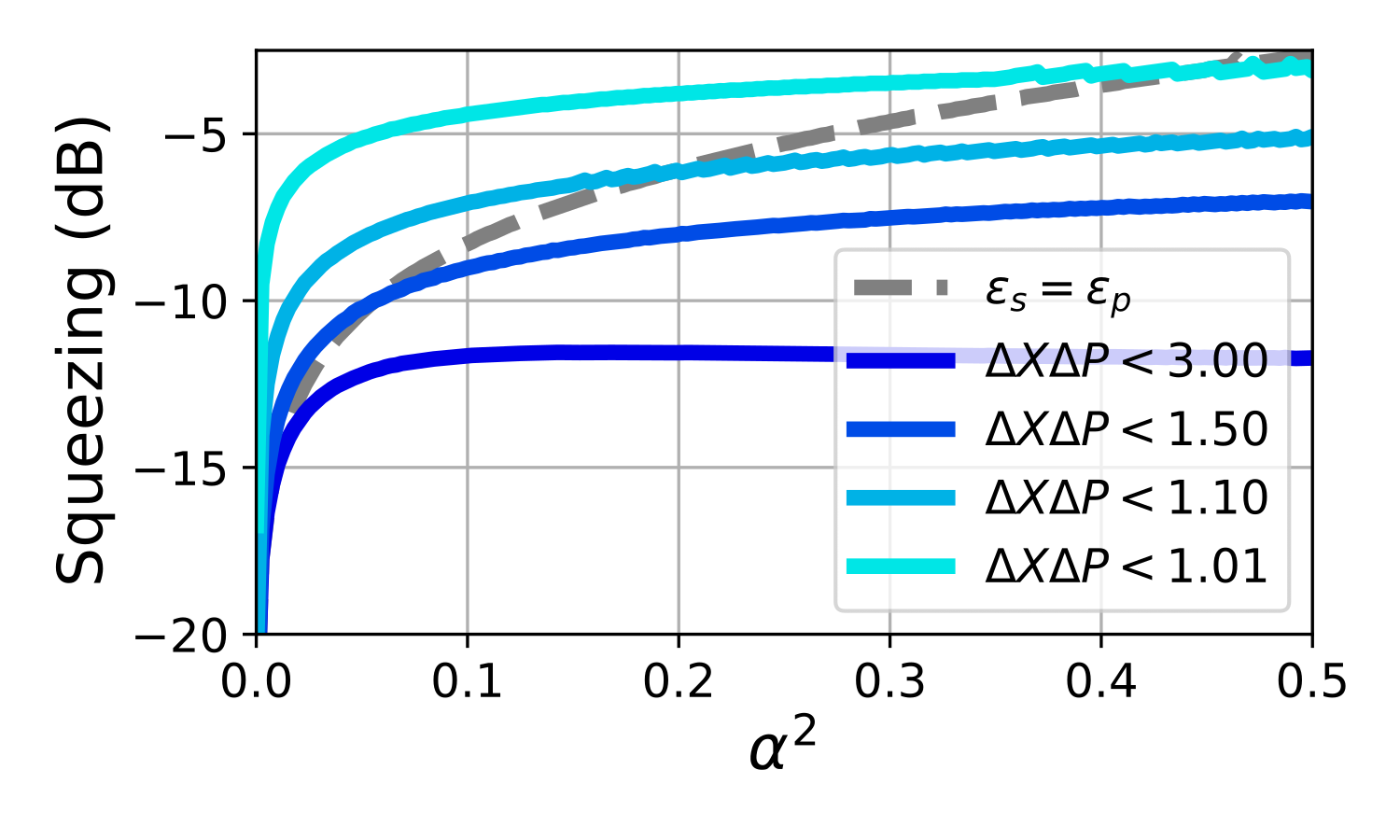}\label{fig:opa-xsqueeze_opt}}
		\caption{The optimal squeezing attainable for a parametric amplifier operating in the \subref{fig:opa-psqueeze_opt} amplifying, phase squeezing and \subref{fig:opa-xsqueeze_opt} deamplifying, amplitude-squeezing regimes constrained to have overall uncertainty below several threshold values. See the details of the dashed $\mathcal{E}_s = \mathcal{E}_p$ curve in the main text.} \label{fig:opa-squeeze_opt}
	\end{figure}
	
	These dynamics result in a complicated parameter dependence of squeezing.  In particular, in both cases, amplitude maxima and squeezing maxima do not coincide, meaning that optimal operation will depend on choosing the interaction duration appropriately. As before, we plot (Fig.~\ref{fig:opa-squeeze_opt}) the squeezing as a function of relative squared output displacement $\alpha^2=\frac{A_s(t)^2}{\mathcal{E}_p^2}$.  In this case we choose the maximum squeezing subject to uncertainty constraints for a given $A_s(t)^2$ for all $t$.  The results are broadly consistent with the parametric oscillator performance; however, while phase-squeezing performance is nearly identical, amplitude squeezing performance is superior when overall uncertainty is allowed to be greater.  In fact, highly squeezed states are possible at high relative brightness; at higher power levels the performance even exceeds that of beam splitter mixing in the previous section.  However, this comes with a caveat: in this regime the required seed \emph{input} brightness is substantial. At high relative output displacement the seed input must be more than an order of magnitude more powerful than the pump input.  The gray dashed line in Fig.~\ref{fig:opa-xsqueeze_opt} shows the cutoff introduced if we restrict the seed input power to that of the pump input power, such that outputs to the right of the line are unavailable with this constraint. So while, in contrast to the parametric oscillator case, the amplifier benefits from significant increases to seed input brightness as a counter to deamplification, in practice the inputs required render the advantage in terms of the relative output power moot.

\section{Optomechanical Squeezer}
The third method that we consider relies on coupling of optical fields to a mechanical resonator. While squeezing has been demonstrated using a wide variety of implementations \cite{Brooks2012opm,SafaviNaeini2013opm,purdy2013opm,Xiong22opm}, here we consider a dissipative optomechanical light squeezing scheme \cite{kronwald2014dissipative}. In this method, the light is squeezed via a strong optomechanical coupling where the mechanical degree of freedom is squeezed due to the interaction with the two-tone optical pumping into the optical cavity. While the authors in \cite{kronwald2014dissipative} considered only the case of vacuum optical seeding (i.e., zero-amplitude seeding into the optical cavity), we expand the analysis to include now the case where the optical cavity is seeded by a non-zero-amplitude coherent light whose frequency is resonant with the optical cavity.  

The optomechanical system's Hamiltonian is given by \cite{aspelmeyer2014cavity}:
\begin{flalign}
	H_\mathrm{OM} = \hbar \omega_\mathrm{cav} a^\dagger a + \hbar \Omega b^\dagger b - \hbar g (b^\dagger + b) a^\dagger a,\label{eq:om-ham}
\end{flalign}
where $a, a^\dagger$ are the annihilation and the creation operators for the photonic field, $b, b^\dagger$ are those for the phononic field, and $\omega_\mathrm{cav}$, $\Omega$ are the resonant frequencies of the photonic and phononic cavities, respectively. Here, $g$ is the single-photon optomechanical coupling coefficient.  The cavity is coupled to the external environment (including the input fields) given by the field operators
	\begin{align}
        a_{\mathrm{in}}&=a_{\mathrm{in}}^0+\expn{i\omega_\mathrm{cav} t}\left(E_p(t)+E_s(t)\right),\nnl
		E_p(t)&=\mathcal{E}_{+} e^{i \Omega t} + \mathcal{E}_- e^{-i \Omega  t},\nnl
		E_s(t)&= \mathcal{E}_0,\label{eq:om-input} 
	\end{align} 
	where $a_{\mathrm{in}}^0$ is a vacuum noise operator and $\mathcal{E}$ is the classical driving field \cite{gardiner1985input}. We set the seed to be $X$ displaced, so that $\mathcal{E}_0^{\rm in}$ is real.  The phononic cavity is coupled to thermal noise $b_{\mathrm{in}}$. 
	
    Assuming that $\Gamma <\kappa\ll\Omega$, generating and solving the equations of motion (Appendix C) we obtain amplitude squeezing for $\mathcal{E}_\pm$ imaginary given by
\begin{flalign}
	\alpha^2&=\frac{\left(1-\mathcal{C}\mathcal{D}\right)^3}{2C^2\left(1+\mathcal{D}^2\right)},\nnl
	\left(\Delta X_s\right)^2&=\frac{\left(1+\mathcal{D}^2\right)\left(1-\mathcal{C}\mathcal{D}\right)}{2}+\mathcal{C}\mathcal{D}^2\left(2\bar{n}+1\right),\nnl
	\left(\Delta P_s\right)^2&=\frac{\left(1-\mathcal{C}\mathcal{D}\right)^2}{\left(1+\mathcal{C}\mathcal{D}\right)^2}+\frac{4}{\left(1+\frac{1}{\mathcal{C}\mathcal{D}}\right)^2}\frac{1}{\mathcal{C}\mathcal{D}^2}\left(2\bar{n}+1\right),\label{eq:om-squeeze}
\end{flalign} 
where $\bar{n}$ is the thermal occupation of the mechanical oscillator, $\mathcal{C}=\frac{4g^2\abs{\mathcal{E}_+-\mathcal{E}_-}^2}{\Gamma\Omega}$ and is determined by the system parameters, $\mathcal{D}=\frac{1}{2}\frac{\abs{\mathcal{E}_++\mathcal{E}_-}}{\abs{\mathcal{E}_+-\mathcal{E}_-}}$ and captures the difference in amplitude of the two probe inputs, and we have defined the output squared displacement ratio to be $\alpha^2=\frac{\mathcal{E}_0^{\rm out}}{\abs{\mathcal{E}_-}^2+\abs{\mathcal{E}_+}^2}$.
When $\mathcal{E}_\pm$ are real, the expressions for $\Delta X_s$ and $\Delta P_s$ are interchanged, yielding phase squeezing. 

The first terms on RHS of the quadrature expressions can be traced to noise contributed by from the probe fields, while the second term originates from the mechanical oscillator.  We note that, for a vacuum input leading to a squeezed vacuum output, $\mathcal{C}\mathcal{D}=1$ so that
\begin{flalign*}
	\alpha^2&=0,\\
	\left(\Delta X_s\right)^2&=\frac{1}{\mathcal{C}}\left(2\bar{n}+1\right),\\
	\left(\Delta P_s\right)^2&=\mathcal{C}\left(2\bar{n}+1\right),
\end{flalign*}
for amplitude squeezing squeezing (phase squeezing of course yields the same result with the expressions for the quadratures interchanged), consistent with \cite{kronwald2014dissipative}. However, allowing even a low amplitude seed ouptut has a dramatic impact on squeezing.  To better understand the relationship between the quadrature variances and the output power, we can write for $\bar{n}=0$ the quadratures to leading order in $\alpha^2$ as
\begin{flalign*}
\left(\Delta X_s\right)^2&\approx\mathcal{D}\left[1+\frac{\left(1-\mathcal{D}\right)^2}{\mathcal{D}}\left(\frac{\left(1+\frac{1}{\mathcal{D}^2}\right)}{4}\alpha^2\right)^{\frac{1}{3}}\right],\\
\left(\Delta P_s\right)^2&\approx\frac{1}{\mathcal{D}}\left[1-\left(1-\mathcal{D}\right)\left(\frac{\left(1+\frac{1}{\mathcal{D}^2}\right)}{4}\alpha^2\right)^{\frac{2}{3}}\right].
\end{flalign*}

	\begin{figure}[!tb]
	\centering
	\includegraphics[width=0.50\textwidth]{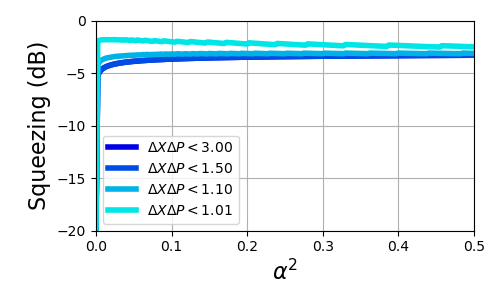}	
	\caption{The optimal squeezing attainable for an optomechanical dissipative squeezer assuming $\bar{n}=0$.  Each curve shows the best squeezing possible while maintaining total uncertainty below a given threshold.} \label{fig:om-squeeze}
\end{figure}
  It is clear that the squeezing is rapidly reduced as the output power increases, as does the deviation of the total uncertainty from unity.  When $\mathcal{D}$ is small, which is the condition that maximizes squeezing in the vacuum output case, the additional amplitude noise increases rapidly with desired output power, an effect that is exacerbated by the cube root applied to the $\alpha^2$.  While the phase noise is somewhat reduced, the scaling of this reduction with seed output power is much weaker. This is shown numerically (using Eq.\eqref{eq:om-squeeze}) in Fig.~\ref{fig:om-squeeze}, which as before shows the squeezing as a function of relative squared output displacement under different overall uncertainty constraints, here generated for different values of $\mathcal{C}$ and $\mathcal{D}$.  For brighter outputs modest squeezing is possible regardless of uncertainty desired, but at low finite outputs the performance is quite poor as demonstrated above, with less squeezing actually possible for low desired uncertainty than at higher powers.  The brightness vs. squeezing tradeoffs for an optomechanical squeezer are severe enough that the optimal strategy is to produce a squeezed vacuum and displace via a beam splitter.

\section{General Case}
All of the methods explored here exhibit trade-offs between displacement and both squeezing and overall uncertainty. While the specific physics involved dictate the form and severity of these tradeoffs, their existence follows from the use of a pump field to both squeeze and displace the output, and we expect the performance of any parametric-amplification-based method to be similarly afflicted.  To see this we note that these Hamiltonians contain cubic terms that yield terms of the form $c_sc_p$, where $c_s$ and $c_p$ may be creation or annihilation operators in the fields, in the equation of motion for the intended squeezed output operator.  Partitioning the fields into amplitudes and noise components and linearizing, the contribution to the noise evolution will have the form $C_sc_p+C_pc_s$, where $C_p$ and $C_s$ are the C-number amplitudes.  The term $C_pc_s$ generates the squeezing while $C_sc_p$ mixes in the noise from the pump field; when the seed amplitude is zero, only squeezing is performed and a pure state results, while nonzero seed amplitude – required to produce displaced output states – necessarily introduces and adds additional noise from the pump field.  Therefore any system with this form of Hamiltonian will be subject to the tradeoffs discussed.
 
\section{Conclusion}
	We presented the performance analysis of generating highly-displaced bright squeezed light from a beam-splitter mixing given a prepared squeezed vacuum as well as generating directly from a seeded optical parametric oscillator, a seeded optical parametric amplifier, and a seeded dissipative optomechanical light squeezer. For each method, we provided a full-quantum-model-based derivation of the expected squeezed (reduced) noise from the main output of the system as a function of mixing degree between the vacuum squeezed mode and a strong coherent-state input. Our results demonstrate that there are limits inherent to the physics of each method that result in tradeoffs between output power (brightness), squeezing, and overall uncertainty that are generic to use of pump field to cause squeezing.  All squeezing methods suffer degradation of both squeezing and total uncertainty as the desired brightness is increased.  Phase squeezing in either parametric oscillators or amplifiers, as well as amplitude squeezing in parametric amplifiers, are able to outperform the beam splitter mixing of a prepared squeezed vacuum with a bright coherent state, while the optomechanical dissipative squeezer and amplitude-squeezing parametric oscillator perform worse.  Additionally, in all cases, it is important to note that parameters that produce the greatest squeezed vacuum do not necessarily provide the best results for bright squeezed states.

	\section*{Acknowledgements}
    The article was supported by the Laboratory Directed Research and Development program at Sandia National Laboratories and has been authored by an employee of National Technology \& Engineering Solutions of Sandia, LLC under Contract No. DE-NA0003525 with the U.S. Department of Energy (DOE). The employee owns all right, title and interest in and to the article and is solely responsible for its contents. The United States Government retains and the publisher, by accepting the article for publication, acknowledges that the United States Government retains a non-exclusive, paid-up, irrevocable, world-wide license to publish or reproduce the published form of this article or allow others to do so, for United States Government purposes. The DOE will provide public access to these results of federally sponsored research in accordance with the DOE Public Access Plan https://www.energy.gov/downloads/doe-public-access-plan This paper describes objective technical results and analysis. Any subjective views or opinions that might be expressed in the paper do not necessarily represent the views of the U.S. Department of Energy or the United States Government.

\appendix
\section{Beamsplitter}
The Heisenberg equations of motion for the beam-splitter Hamiltonian (Eq.\eqref{eq:bs-ham}) in the rotating frame are simply
\begin{flalign*}
	\dot{a_s}=i\eta a_s,\\
	\dot{a_p}=i\eta a_p,
\end{flalign*}
which can be straightforwardly solved to yield
\begin{flalign*}
	a_p(t)=a_p(0) \cos \eta t - i a_p(0) \sin \eta t.
\end{flalign*}
Based on the initial state described above, we have
\begin{flalign*}	
	a_s(0)&=S_s (B)^\dagger a_s S_s (B)= a_s \cosh r - e^{i \gamma } a_s^\dagger \sinh r, \nonumber \\
	a_p(0)&=D_p (\mathcal{E})^\dagger a_p D_p (\mathcal{E}) = a_p + \mathcal{E},
\end{flalign*}
so that we can compute the output variances (for $B$ and $\mathcal{E}$ real) as 
\begin{align*}
	\left(\Delta X_s\right)^2 &= \bra{\psi_o}_a X_s^2 \ket{\psi_o}_s -\left( \bra{\psi_o}_s X_s \ket{\psi_o}_s \right)^2\\
	&=e^{-2 B} \cos^2 (\eta t) + \sin^2 (\eta t),\\
	\left(\Delta P_s\right)^2 &=\bra{\psi_o}_a P_s^2 \ket{\psi_o}_s -\left( \bra{\psi_o}_s P_s \ket{\psi_o}_s \right)^2\\
	&=e^{2 B} \cos^2 (\eta t) + \sin^2 (\eta t).\\
\end{align*}
This gives the squeezing, which is in $X$ for $B>0$ and in $P$ for $B<0$. The overall uncertainty is then
\begin{flalign}
	\Delta X_s\Delta P_s=\sqrt{1+2\cos^2 (\eta t) \sin^2 (\eta t)\left(\cosh 2B -1\right)}.
\end{flalign}  

The relative squared displacement of the output signal is computed as 
\begin{align}
	\alpha^2 &= \frac{\bra{\psi_o} X_s \ket{\psi_o}^2 + \bra{\psi_o} P_s \ket{\psi_o} }{4\abs{\mathcal{E}}^2}^2\nnl
	&=\frac{\abs{\mathcal{E}}^2\sin^2 (\eta t)}{\abs{\mathcal{E}}^2}\nnl
	&=\sin^2 (\eta t)
\end{align}

\section{Parametric Oscillator}
	The rotating frame equations of motion for a parametric oscillator where input and output occur on a single side of the cavity, using the Hamiltonian in equation \eqref{eq:OPO-hamiltonian}, are given as
\begin{align}
	\dot{a_s} &= g a_s^\dagger a_p -\frac{\kappa}{2} a_s -\sqrt{\kappa}a_s^{\rm in}, \nonumber \\
	\dot{a_p} &= - (g/2) a_s^2 -\frac{\kappa}{2} a_p -\sqrt{\kappa}a_p^{\rm in}, \nnl
	a_s^{\rm out}(t)&=\sqrt{\kappa}a_s(t)+a_s^{\rm in}, \nnl
	a_p^{\rm out}(t)&=\sqrt{\kappa}a_p(t)+a_p^{\rm in},
\end{align} 
where $a_s$ and $a_p$ are intracavity operators, $a_s^{\rm in/out}$ and $a_p^{\rm in/out}$ are the input/output field operators, and $\mathcal{E}_s^{\rm in}$ and $\mathcal{E}_p^{\rm in}$ are the input field displacements. Here, $\kappa$ is the cavity photon decay rate. 

Assuming steady state operation, by linearizing the above equations by letting $a_p\rightarrow A_p+a_p$, $a_s\rightarrow A_s+a_s$, where $A_p=\langle a_p\rangle$ and $A_s=\langle a_s\rangle$, and dropping terms of order greater than one in the new field operators, we obtain
\begin{align}
	A_s &= \frac{2g}{\kappa} A_s^* A_p  -\frac{2}{\sqrt{\kappa}}\mathcal{E}_s^{\rm in}, \nonumber \\
	A_p &= - \frac{g}{\kappa} A_s^2  -\frac{2}{\sqrt{\kappa}}\mathcal{E}_p^{\rm in}, \nnl
	\mathcal{E}_s^{\rm out}&=\sqrt{\kappa}A_s+\mathcal{E}_s^{\rm in}, \nnl
	\dot{a_s} &= g\left( a_s^\dagger A_p+A_s^* a_p\right) -\frac{\kappa}{2} a_s -\sqrt{\kappa}a_s^{\rm in}, \nonumber \\
	\dot{a_p} &= - g A_sa_s -\frac{\kappa}{2} a_p -\sqrt{\kappa}a_p^{\rm in}, \nnl
	a_s^{\rm out}(t)&=\sqrt{\kappa}a_s(t)+a_s^{\rm in}, 
\end{align} 
where $\mathcal{E}_p^{\rm in}$ and $\mathcal{E}_s^{\rm in}$ are input field amplitudes and $\mathcal{E}_s^{\rm out}$ is the squeezed output amplitude. In frequency space the field equations of motion become
\begin{flalign*}
	a_s(\omega)&=g\left(a_s(-\omega)\dg A_p+A_s^*a_p(\omega)\right)-\sqrt{\kappa}a_s^{\rm in}(\omega), \\
	a_p(\omega)&=-gA_sa_s(\omega)-\sqrt{\kappa}a_p^{\rm in}(\omega).
\end{flalign*}
Solving for $A_s$,$A_p$ and $a_s(0)$ as in \cite{lariontsev2002characteristics}, we find
\begin{flalign*}
	\chi&=g\mathcal{E}_s^{\rm in}\left(1-\sqrt{1+\frac{1}{2g^2(\mathcal{E}_s^{\rm in})^2}\left(\frac{4\sqrt{\kappa}g\mathcal{E}_p^{\rm in}+\kappa^2}{3\kappa}\right)^3}\right), \\
	A_s&=\frac{-\sqrt{\kappa}\chi^{1/3}}{g}+\frac{4g\mathcal{E}_p^{\rm in}+\kappa^{3/2}}{6g\chi^{1/3}}, \\
	A_p&=-\frac{gA_s^2}{\kappa}-\frac{2}{\sqrt{\kappa}}\mathcal{E}_p^{\rm in},\\
	a_s(0)&=-\sqrt{\kappa}\frac{\left(g^2A_s^2+\frac{\kappa^2}{4}\right)\left[\frac{\kappa}{2}a_s^{\rm in}(0)+gA_sa_p^{\rm in}(0)\right]+\frac{\kappa gA_p}{2}\left[\frac{\kappa}{2}a_s^{\rm in\dagger}(0)+gA_sa_p^{\rm in\dagger}(0)\right]}{\left(g^2\abs{A_s}^2+\frac{\kappa^2}{4}\right)^2-\left(\frac{\kappa gA_p}{2}\right)^2}
\end{flalign*}
where we have taken $A_p$ and $A_s$ to be real. This yields the output quadratures 
\begin{flalign}
	\left(\Delta X_s\right)^2&=\frac{\left(g^2A_s^2-\frac{\kappa gA_p}{2}-\frac{\kappa^2}{4}\right)^2+\left(\kappa g A_s\right)^2}{\left(g^2A_s^2-\frac{\kappa gA_p}{2}+\frac{\kappa^2}{4}\right)^2}, \\
	\left(\Delta P_s\right)^2&=\frac{\left(g^2A_s^2+\frac{\kappa gA_p}{2}-\frac{\kappa^2}{4}\right)^2+\left(\kappa g A_s\right)^2}{\left(g^2A_s^2+\frac{\kappa gA_p}{2}+\frac{\kappa^2}{4}\right)^2},\label{eq:opo-quad}
\end{flalign}
and relative squared output displacement
\begin{flalign*}
	\alpha^2=\left(\frac{\mathcal{E}_s^{\rm out}}{\mathcal{E}_p^{\rm in}}\right)^2=\left(\frac{\mathcal{E}_s^{\rm in}+\sqrt{\kappa}A_s}{\mathcal{E}_p^{\rm in}}\right)^2.\label{eq:opo-output}
\end{flalign*}

\section{Parametric Amplifier}
 The equations of motion for the parametric amplifier are 
\begin{align}
	\dot{a_s} &= g a_s^\dagger a_p, \nonumber \\
	\dot{a_p} &= - (g/2) a_s^2,
\end{align}
with an initial condition of
\begin{equation}
	\ket{\psi(0)} = D_s (\mathcal{E}_s) \ket{0}_s \otimes D_p (\mathcal{E}_p) \ket{0}_p,
\end{equation}
where initially the signal is seeded with a coherent state with an amplitude $\mathcal{E}_s$ and the pump's initial condition is also a coherent state with an amplitude $\mathcal{E}_p$. Here, $D_{s,p} (\mathcal{E})$ is a displacement operator as in equation \eqref{eq:displacement}. 
Once again letting $a_p\rightarrow A_p+a_p$, $a_s\rightarrow A_s+a_s$, where $A_p=\langle a_p\rangle$ and $A_s=\langle a_s\rangle$,
\begin{flalign}
	\dot{A_s} &= g A_s^* A_p, \nnl
	\dot{A_p} &= - \frac{g}{2} A_s^2,  \nnl
	\dot{a_s} &= g\left( a_s^\dagger A_p+A_s^* a_p+a_s\dg a_p\right),  \nonumber \\
	\dot{a_p} &= - \frac{g}{2} \left(2A_sa_s -a_s^2\right), \nnl
	A_s(0) &= \mathcal{E}_s, \nnl
	A_p(0) &= \mathcal{E}_p. \label{eq:opa_eom}
\end{flalign}

It is important to emphasize that the pump that drives the parametric process cannot be assumed to be a constant as before and in previous analyses \cite{milburn1981production}. 
To see this we can analytically solve for $A_s(t)$ and $A_p(t)$, yielding 
\begin{align}
	A_s(t) &= \left( 2 c_1 (1 - \tanh^2(g t \sqrt{c_1} + \sqrt{c_1} c_2) \right)^{1/2}, \nonumber \\
	A_p (t) &= - \sqrt{c_1} \tanh(gt \sqrt{c_1} + \sqrt{c_1} c_2),
\end{align}
where
\begin{align}
	c_1 &= \mathcal{E}_p^2 + \mathcal{E}_s^2/2, \nonumber \\
	c_2 &= - \tanh^{-1} \left(\mathcal{E}_p/\sqrt{\mathcal{E}_p^2 + \mathcal{E}_s^2/2} \right)/\sqrt{\mathcal{E}_p^2 + \mathcal{E}_s^2/2}.
\end{align}
\section{Optomechanical Squeezer}

Starting from the optomechanical system's Hamiltonian and input field descriptions (Eqs. \eqref{eq:om-ham} and \eqref{eq:om-input}), the equations of motion in the rotating frame are 
\begin{flalign*}
	\dot{a}&=ig\left(b\dg\expn{i\Omega t}+b\expn{-i\Omega t}\right)a
	-\frac{\kappa}{2}a-\sqrt{\kappa}\left(a^{in}+\mathcal{E}_p(t)+\mathcal{E}_s(t)\right),\\
	\dot{b}&=iga\dg a\expn{i\Omega t}-\frac{\Gamma}{2}-\sqrt{\Gamma}b^{in}.
\end{flalign*}
which in Fourier space become 
\begin{widetext}
	\begin{flalign*}
		a(\omega)&=\frac{ig\int d\omega' \left[a(\omega')b(-\omega+\omega'+\Omega)\dg+a(\omega')b(\omega-\omega'+\Omega)\right]-\sqrt{\kappa}\left(a^{in}(\omega)+\mathcal{E}_p(\omega)+\mathcal{E}_s(\omega)\right)}{\frac{\kappa}{2}-i\omega},\\
		b(\omega)&=\frac{ig\int d\omega' a(-\omega')\dg a(\omega-\omega'-\Omega)-\sqrt{\Gamma}b^{in}(\omega)}{\frac{\Gamma}{2}-i\omega},
	\end{flalign*}
	we take $a(\omega)\rightarrow A(\omega)+a(\omega)$ and $b(\omega)\rightarrow b(\omega)+b(\omega)$ and discard the terms higher than linear order in the field operators to obtain
	\begin{flalign*}
		A(\omega)&=\frac{ig\int d\omega' A(\omega')\left[B(-\omega+\omega'+\Omega)\dg+B(\omega-\omega'+\Omega)\right]-\sqrt{\kappa}\left(\mathcal{E}_p(\omega)+\mathcal{E}_s(\omega)\right)}{\frac{\kappa}{2}-i\omega}, \\
		B(\omega)&=\frac{ig\int d\omega' A(-\omega')\dg A(\omega-\omega'-\Omega)}{\frac{\Gamma}{2}-i\omega}, \\
		a(\omega)&=\frac{1}{\frac{\kappa}{2}-i\omega}\left\{ig\int d\omega' \left[A(\omega')\left(b(-\omega+\omega'+\Omega)\dg+b(\omega-\omega'+\Omega)\right)\right.\right. \\
		&\left.\left.+a(\omega')\left(B(-\omega+\omega'+\Omega)\dg+B(\omega-\omega'+\Omega)\right)\right]-\sqrt{\kappa}a^{in}(\omega)\right\},\\
		b(\omega)&=\frac{ig\int d\omega'\left[ A(-\omega')\dg a(\omega-\omega'-\Omega)+ a(-\omega')\dg A(\omega-\omega'-\Omega)\right]-\sqrt{\Gamma}b^{in}(\omega)}{\frac{\Gamma}{2}-i\omega}.
	\end{flalign*}
	The driving fields will ensure that at steady state the average amplitudes will be delta functions at multiples of $\Omega$; \ie we can write $A(\omega)=\sum_n A_{n}\delta(\omega-n\Omega)$ and $B(\omega)=\sum_n B_{n}\delta(\omega-n\Omega)$. We then have
	\begin{flalign*}
		A_0&=\frac{ig\sum_{j} A_j\left[B_{1+j}\dg+B_{1-j}\right]-\sqrt{\kappa}\mathcal{E}_0}{\frac{\kappa}{2}},\\
		A_{+1}&=\frac{ig\sum_{j} A_j\left[B_{j}\dg+B_{2-j}\right]-\sqrt{\kappa}\mathcal{E}_+}{\frac{\kappa}{2}-i\Omega},\\
		A_{-1}&=\frac{ig\sum_{j} A_j\left[B_{2+j}\dg+B_{-j}\right]-\sqrt{\kappa}\mathcal{E}_-}{\frac{\kappa}{2}+i\Omega},\\
		B_0&=\frac{igA_{-j}\dg A_{-j-1}}{\frac{\Gamma}{2}},\\
		a(\omega)&=\frac{1}{\frac{\kappa}{2}-i\omega}\left\{ig \left[A_j\left(b(-\omega+(1+j)\Omega)\dg+b(\omega+(1-j)\Omega)\right)+\right.\right.\\
		&\left.\left.a(\omega+(j-1)\Omega)B_j\dg+a(\omega+(1-j)\Omega)B_j\right]-\sqrt{\kappa}a^{in}(\omega)\right\},\\
		b(\omega)&=\frac{ig\int d\omega'\left[ A_{-j}\dg a(\omega-(j+1)\Omega)+ a(\omega-\Omega)\dg A_j\right]-\sqrt{\Gamma}b^{in}(\omega)}{\frac{\Gamma}{2}-i\omega}.
	\end{flalign*}
	
	We will assume that $\kappa\ll\Omega$ so that we can restrict our consideration to $A_0$ and $A_{\pm\Omega}$, the latter of which we will write as $A_{\pm}$ for brevity.  As a result, since ultimately we desire $a(0)$, upon linearizing we need only concern ourselves with the noise operator at $\omega=0$ and $\omega=\pm\Omega$ and write them as $a_0$ and $a_\pm$, respectively.  We will also assume $\Gamma \ll \Omega$ so that we need only consider $b_0=b(0)$. We then have
	
	\begin{flalign*}
		A_0&=\frac{1}{\sqrt{\kappa}}\frac{1}{\frac{1}{2}+\frac{g^2}{\kappa\Gamma}\left(\abs{A_+}^2-\abs{A_-}^2\right)}\mathcal{E}_0,\\
		A_\pm&=\frac{\sqrt{\kappa}}{\left(\frac{\kappa}{2}\mp i\Omega\right)^2}\left[\mp\frac{2g^2A_0^2}{\Gamma}\mathcal{E}_\mp^*-\left(\frac{\kappa}{2}\mp i\Omega\pm \frac{2g^2\abs{A_0}^2}{\Gamma}\right)\mathcal{E}_\pm\right],\\
		a_{0}&=\frac{g^2A_{0}\left[A_{-}\dg  a_{-}-A_{+}\dg a_{+}+A_{-} a_{-}\dg-A_{+}a_{+}\dg\right]-ig\sqrt{\Gamma}\left[A_{-}b_{0}^{in,\dagger}+A_+b_{0}^{in}\right]-\frac{\sqrt{\kappa}\Gamma}{2}a^{in}_{0}}{\frac{\kappa\Gamma}{4}+g^2\left(\abs{A_+}^2-\abs{A_-}^2\right)},\\
		a_{\pm}&\approx\frac{\pm g^2\left[\left(2A_{\mp}\dg A_{0}+A_0\dg A_{\pm} \right)a_{0}+A_0 A_{\pm}a_0\dg+A_0 A_{0}a_{\mp}\dg\right]-ig\sqrt{\Gamma}A_0b_{0}^{in,\pm\dagger}-\frac{\sqrt{\kappa}\Gamma}{2} a^{in}_{\pm}}{\frac{\kappa\Gamma}{4}\mp i\frac{\Gamma\Omega}{2}\mp g^2\abs{A_0}^2}.
	\end{flalign*}

	Then, assuming $A_0$ is real, solving these equations we find that the output quadrature
	\begin{flalign*}
		\left(\Delta X_s\right)^2&=\abs{\sqrt{\kappa}x_0+1}^2+\kappa\abs{x_+}^2+\kappa\abs{x_-}^2+\kappa\abs{x_b}^2,\\
		x_0&=\frac{\left[i\frac{2C_g^4A_0^2\left(C_g^2A_0^2-C_\Omega^2\Im\left[A_+A_-\right]\right) }{C_\Omega^6\abs{A_-^*+A_+}^4}+\frac{1}{2}\left(1+\Delta\right)\left(\Delta+2\frac{C_g^4A_0^2}{C_\Omega^4\abs{A_-^*+A_+}^2}\right)\right]}{\left(1+\Delta\right)\left(\frac{1}{2}\left(1+\Delta\right)+\frac{C_g^4A_0^2}{C_\Omega^4\abs{A_-^*+A_+}^2}\right)},\\
		x_\pm&=\frac{1}{\sqrt{\kappa}}\frac{-C_g^2\frac{A_0}{\abs{A_-^*+A_+}}\left(C_g^2\frac{A_0^2}{\abs{A_-^*+A_+}^2}\frac{A_\mp+A_\pm^*}{\abs{A_-^*+A_+}}-i \frac{A_\pm^*}{\abs{A_-^*+A_+}} C_\Omega^2\right)}{C_\Omega^4\left(\frac{1}{2}\left(1+\Delta\right)+\frac{C_g^4A_0^2}{C_\Omega^4\abs{A_-^*+A_+}^2}\right)},\\
		x_b&=\frac{1}{\sqrt{\Gamma}}\frac{i2C_g\left[i2\frac{C_g^2}{C_\Omega^2}\frac{A_0^2}{\abs{A_-^*+A_+}^2}\frac{A_-^*+A_+}{\abs{A_-^*+A_+}}\left(\frac{C_g^2A_0^2\left(C_g^2A_0^2-C_\Omega^2\Im\left[A_+A_-\right]\right)}{C_\Omega^4\abs{A_-^*+A_+}^4}+\frac{1+\Delta}{2}\right)+\frac{1+\Delta}{2}\frac{A_-^*-A_+}{\abs{A_-^*+A_+}}\right]}{\left(1+\Delta\right)\left(\frac{1}{2}\left(1+\Delta\right)+\frac{C_g^4A_0^2}{C_\Omega^4\abs{A_-^*+A_+}^2}\right)},\\
		\left(\Delta P_s\right)^2&=\abs{\sqrt{\kappa}p_0+1}^2+\kappa\abs{p_+}^2+\kappa\abs{p_-}^2+\kappa\abs{p_b}^2,\\
		p_0&=-\frac{1}{\sqrt{\kappa}\left(1+\Delta\right)},\\
		p_\pm&=0,\\
		p_b&=\frac{2C_g}{\sqrt{\Gamma}\left(1+\Delta\right)},\\
        \alpha^2&=\frac{\Delta^2}{\left(1+\Delta\right)}\frac{\mathcal{E}_{0}^2}{\abs{\mathcal{E}_{+}}^2+\abs{\mathcal{E}_{-}}^2},
	\end{flalign*}
\end{widetext}
where we have defined the quantities
\begin{flalign*}
	C_g&=\frac{g\abs{A_++A_-^*}}{\kappa},\\
	C_\Omega^2&=\frac{\Gamma\Omega}{2\kappa^2},
\end{flalign*}
for convenience.
The variance can be nearly -- though not quite exactly -- minimized by	setting $\Delta=-2\frac{C_g^4A_0^2}{C_\Omega^4\abs{A_-^*+A_+}^2}$ and taking $\mathcal{E}_\pm$ to be purely imaginary and of opposite sign, resulting in Eqs.\eqref{eq:om-squeeze}
\bibliography{ref}

\end{document}